\documentclass[twocolumn,preprintnumbers]{revtex4-1}
\usepackage{amsmath}
\usepackage{dcolumn}
\usepackage{bm}
\usepackage{epsfig}
\usepackage{epstopdf}
\usepackage{graphicx}
\usepackage{amsfonts}
\usepackage{amssymb}
\usepackage{mathrsfs}
\usepackage{color}
\usepackage{multirow}
\usepackage{appendix}
\usepackage{float}
\usepackage{graphicx,amsmath}
\usepackage{color,xcolor}
\usepackage[bookmarks=false]{hyperref}
\hypersetup{colorlinks=true,citecolor=blue,linkcolor=blue,urlcolor=blue,pdfstartview=FitH,bookmarksopen=true}
\begin{document}
%\linenumbers
%\pagewiselinenumbers
%\switchlinenumbers

\title{Experimental Verification of Demon-Involved Fluctuation Theorems}
\author{L.-L. Yan$^{1,2}$}
\email[These authors contributed equally]{}
\author{J.-T. Bu$^{3,4}$}
\email[These authors contributed equally]{}
\author{Q. Zeng$^{5}$}
\email[These authors contributed equally]{}
\author{K. Zhang$^{6,7}$}
\author{K.-F. Cui$^{1}$}
\author{F. Zhou$^{3}$}
\email{zhoufei@wipm.ac.cn}
\author{S.-L. Su$^{1,2}$}
\author{L. Chen$^{3,8}$}
\author{J. Wang$^{9,10}$}
\email{jin.wang.1@stonybrook.edu}
\author{Gang Chen$^{1,2}$}
%\email{chengang971@163.com}
\author{M. Feng$^{3,8}$}
\email{mangfeng@wipm.ac.cn}
\affiliation{$^{1}$Key Laboratory of Materials Physics, Ministry of Education, School of Physics and Laboratory of Zhongyuan Light, Zhengzhou University, Zhengzhou 450001, China \\
$^{2}$Institute of Quantum Materials and Physics, Henan Academy of Sciences, Zhengzhou 450046, China \\
$^{3}$State Key Laboratory of Magnetic Resonance and Atomic and Molecular Physics, Wuhan Institute of Physics and Mathematics, Innovation Academy of Precision Measurement Science and Technology, Chinese Academy of Sciences, Wuhan 430071, China\\
$^{4}$School of Physics, University of the Chinese Academy of Sciences, Beijing 100049, China\\
$^{5}$State Key Laboratory of Electroanalytical Chemistry, Changchun Institute of Applied Chemistry, Changchun 130022, China \\
$^{6}$School of Physics, Northwest University, Xi'an 710127, China \\
$^7$Peng Huanwu Center for Fundamental Theory, Xi'an 710127, China \\
$^{8}$Research Center for Quantum Precision Measurement, Guangzhou Institute of Industry Co. Ltd, Guangzhou 511458, China \\
$^{9}$Departments of Physics and Astronomy, State University of New York, Stony Brook, NY 11794-3400, USA \\
$^{10}$Center for Theoretical Interdisciplinary Sciences, Wenzhou Institute, University of Chinese Academy of Sciences, Wenzhou, Zhejiang 325001, China}

\begin{abstract}
The limit of energy saving in the control of small systems has recently attracted much interest due to the concept refinement of the Maxwell demon. Inspired by a newly proposed set of fluctuation theorems, we report the first experimental verification of these equalities and inequalities in a ultracold $^{40}$Ca$^{+}$  ion system, confirming the intrinsic nonequilibrium in the system due to involvement of the demon. Based on elaborately designed demon-involved control protocols, such as the Szilard engine protocol, we provide experimentally quantitative evidence of the dissipative information, and observe tighter bounds of both the extracted work and the demon's efficacy than the limits predicted by the Sagawa-Ueda theorem. Our results substantiate a close connection between the physical nature of information and nonequilibrium processes at the microscale, which help further understanding the thermodynamic characteristics of information and the optimal design of nanoscale and smaller systems.
\end{abstract}

\maketitle

The control of nanoscale and smaller systems is a rapidly developing field in physics, chemistry, and life sciences. Researchers hope to achieve higher control goals at lower costs, such as by fully utilizing observational data to extract more work from the molecular motors~\cite{new1,new2,new3}, or by rewriting information in chip storage units more efficiently with less energy~\cite{LP,nature483-187,njp16-103011,ion-thermo1}. However, the energy consumption and energy utilization efficiency in control are always hard to exactly analyze due to the fundamental limitation imposed by the second law of thermodynamics and Maxwell's demon~\cite{demon1,demon2,demon3,entropy-yan,np11-131,np15-1232,prr2-033082}. Besides, in the process of energy conversion and information processing, certain energy is always dissipated from the system to the environment in the form of heat or entropy, which is called entropy production~\cite{rmp93-035008}. Due to the second law of thermodynamics, entropy production should be non-negative. At the microscopic scale, however, the system fluctuates and the observed entropy production become stochastic~\cite{prl120-010601,prl124-110604,np16-15,ion-thermo2}, which can violate the thermodynamic second law. Nevertheless, the fluctuation theorem indicates that the average of the observed entropy production must satisfy the thermodynamic second law~\cite{fluc1,rpp75-126001,pre60-2721,jsp98-77,Jarzyn}, sometimes also called the generalized second law of thermodynamics~\cite{conn2}.

\begin{figure*}[htpb]
 	\centering
\includegraphics[width=17.5cm,height=3.8cm]{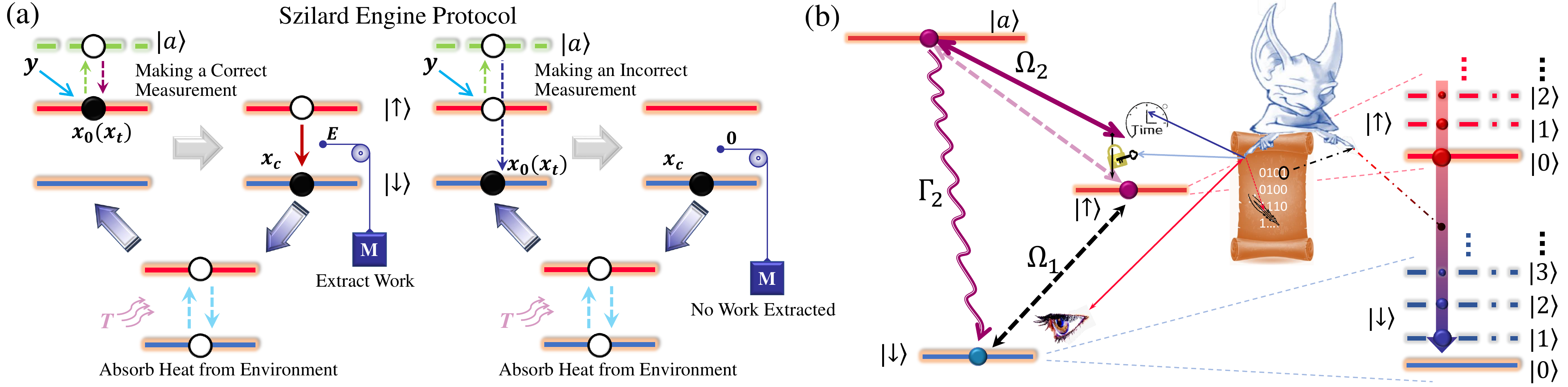}
\includegraphics[width=4.7cm,height=3.7cm]{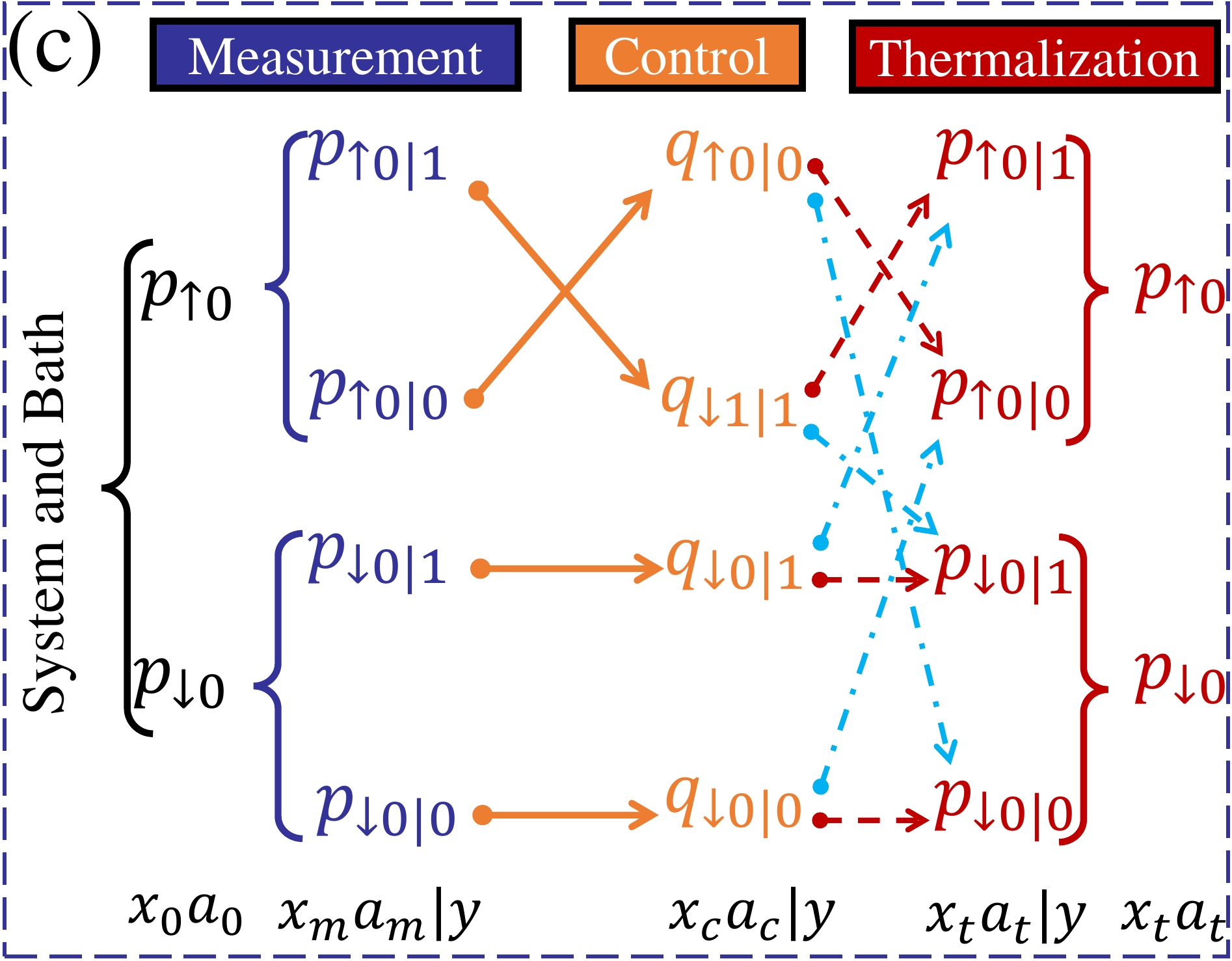}\includegraphics[width=12.8cm,height=3.8cm]{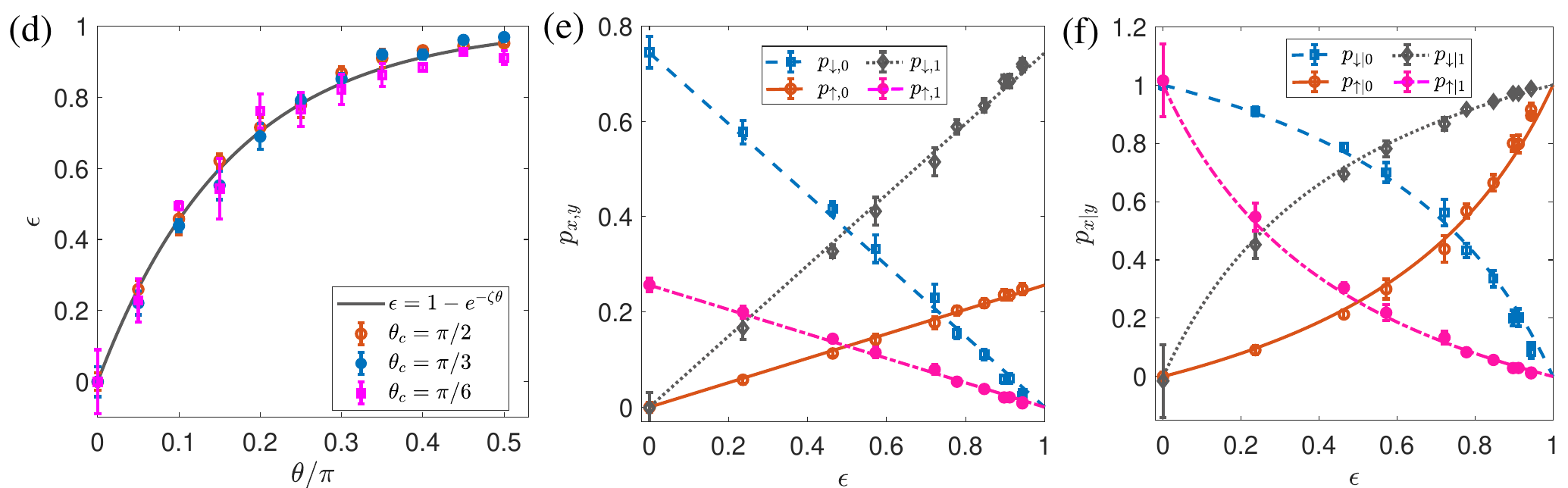}
\caption{Schematic for verifying the demon-involved fluctuation theorems. (a) The Szilard engine feedback control protocol, where the demon (denoted by $y$) makes a noisy measurement on the system (black circle) at the initial time, where the short-lived state $|a\rangle$ (green dashed line) is introduced to simulate measurement noise (as explained in Figure 1(b) and its caption). The demon can successfully extract work of E from the system only if it makes a correct measurement when the system is in the higher energy level (left). In all other cases, including a wrong measurement or a lower energy level (right), the demon extracts no work.
%If the demon makes a correct measurement when the system is in the higher energy level, the demon then extracts work of $E$ from the system, while extracts zero work if the demon makes a wrong measurement.
After the work extraction, the demon couples the system to the environmental heat bath with temperature $T$ to recharge the system for the next cycle. (b)  Experimental scheme based on a trapped $^{40}$Ca$^{+}$ ion, where the system is a qubit encoded by two energy levels of electronic states, i.e., the ground state $\mid\downarrow\rangle=|4S_{1/2},m_J=1/2\rangle$ and the metastable state $\mid\uparrow\rangle=|3D_{5/2},m_J=+3/2\rangle$ with $m_J$ being the magnetic quantum number. An equilibrium Boltzmann distribution and thermalization process of the qubit system is generated by combining the irradiation of 729-nm laser pulse (Rabi frequency $\Omega_1$) with a dephasing process of system. A short-lived state $|a\rangle=|3P_{3/2},m_J=+3/2\rangle$ and a 854-nm laser pulse with Rabi frequence $\Omega_2$ are used for mimicking the noisy measurement by controlling the unlocking and duration of laser. The demon adds a phonon into the vibrational mode through a red-sideband transition carried by the 729-nm laser irradiation, depending on the measurement result "1", i.e., in the upper state $\mid\uparrow\rangle$, where the auxiliary vibrational mode works for storing the energy due to the work output from the system. %The only transition from $\mid\uparrow\rangle|0\rangle$ (the big red dot) to $\mid\downarrow\rangle |1\rangle$ (the big blue dot) exists in the ideal situation, and some other transitions are also available experimentally due to errors and other imperfection.
(c) The probability flow in the Szilard engine controlled by the demon, where the abbreviation $p_{x m|n}:=p(x_0=x, a_{0}=m | y=n)$ with $x=\downarrow, \uparrow$, $m=0, 1, 2, \cdots$, and $n=0,1$, for example, $p_{\uparrow 0|1}$ means that the system is initially in the spin state $\mid\uparrow\rangle$ and the vibrational state $|0\rangle$, and the measurement result is "1"~\cite{SM}. (d) The measurement error under different pulse lengths of the 854-nm laser in the generation of the mixed state $\rho=p_{\downarrow}\mid\downarrow\rangle\langle\downarrow\mid+p_{\uparrow}\mid\uparrow\rangle\langle\uparrow\mid$ with $p_{\downarrow}=(1+\cos\theta_c)/2$ and $p_{\uparrow}=1-p_{\downarrow}$. (e) and (f) The joint and conditional probabilities for $\theta_c=\pi/3$ after the measurement. }
\label{Fig1}
\end{figure*}

However, a system with external control, such as the Szilard engine~\cite{Szilard,prl106-070401,pnas111-13786}, often has an improper entropy production~\cite{SU1,SU2,disobey1,disobey2,disobey3,disobey4,disobey5,disobey6,prl121-210603}, which violates the fluctuation theorem. The improper entropy production was once thought to be caused by Maxwell's demon, or the external control~\cite{demon1,demon2,demon3}. However, together with the mutual information between the demon and the controlled system, the improper entropy production has been shown to satisfy the fluctuation theorem, as given by the Sagawa-Ueda theorem (SUT)~\cite{SU1,SU2,disobey1,disobey2,disobey3,disobey4,disobey5,disobey6}. Nevertheless, SUT only predicts that the work extracted from a system by a demon is not greater than the amount of information gained by the demon's measurement. Recently, a new fluctuation theorem predicts that when a demon manipulates the system, a portion of the information will be dissipated into the environment in the form of heat, called as dissipative information, which makes the actual work obtained be even less~\cite{sa7-eabf1807,prxq_zhang}.  Therefore, it is necessary to consider the dissipative information when designing control protocols in real experiments.

Here we report an experimental verification of the new fluctuation theorems of different entropy productions and dissipative information in a demon-controlled system, as proposed in~\cite{sa7-eabf1807}. Our operations are performed on the $^{40}$Ca$^{+}$ ion confined in a linear Paul trap~\cite{entropy-yan,ion-thermo2}, focusing on the fluctuation theorems related to the demon-involved entropy productions and the corresponding upper bounds of the extracted work from the system as well as the control efficacy of the demon~\cite{sa7-eabf1807}. This constitutes the first experimental evidence of the dissipative information as well as tighter bounds on the extracted work than that from the SUT, confirming the necessity of obeying the second law of thermodynamics. Interestingly, we find that the system can be in a seemingly equilibrium state with vanishing entropy production but still suffers from nonequilibrium energy dissipation, as quantified by nonzero dissipative information, due to its interaction with the demon.

We design in Fig.~\ref{Fig1}(a) a Maxwell's demon based on the Szilard engine control protocol \cite{Szilard,prl106-070401,pnas111-13786} to extract work from a two-level system. The demon performs classical measurements and feedbacks control operations in a cyclic manner, meaning that the system can only be in one of the two energy levels at any given time, e.g., in state $x$. At the initial time of each cycle, the system is in state $x_0$. The demon first performs a classical measurement on the system with the measurement results $y=0\text{ or }1$, where $0$ and $1$ correspond to the lower level ($\mid\downarrow\rangle$) and higher level ($\mid\uparrow\rangle$). Then, the demon takes the control protocol based on the measurement result as follows: $(1)$ If the system is measured to be in the higher level, i.e., $y=1$, the demon executes a feedback control to extract work from the system; $(2)$ If the system is measured to be in the lower energy level, i.e., $y=0$, the demon does nothing, and the system releases no energy.
After the control is finished, the state of the system changes to $x_c$, and the control is performed instantaneously with no heat exchanged. Then, the demon couples the system to the heat bath with temperature $T$ until the end of the cycle, during which the system absorbs heat from the heat bath and finally reaches the equilibrium state $x_t$ (it is also the initial state $x_0$ of the system in the next cycle) with heat bath. Thus, both the initial and the final probability distributions of the system state are given by the Boltzmann distribution $p^{eq}_{x_0}$, where $p^{eq}_{\downarrow}\equiv p_\downarrow=1/(1+e^{-\beta E})$ and $p^{eq}_{\uparrow}\equiv p_\uparrow=1/(1+e^{\beta E})$ in units of $k_B=E=1$.
Besides, if the demon's measurement is noisy that makes the measurement value $y$ not always be equal to the system's initial state $x_0$, the error probability of the measurement, given by the conditional probability $\epsilon \equiv p(y\neq x_0|x_0)$, leads to probabilistic errors in the demon's control of the system, causing the dissipative information. In Fig.~\ref{Fig1}(c), we give the probability flow of complete measurement-control-thermalization for the Szilard engine protocol.

Before unfolding the experimental details, we briefly elucidate the demon-involved fluctuation theorems~\cite{sa7-eabf1807}. The conditional entropy production, denoted by $\sigma_{X|Y}$ (where $X$ represents the system and $Y$ represents the demon), fully contains the information about the demon's control of the system, which is the real entropy generation of system and indicates that the system's entropy production depends on the demon's control. If the details of the demon's control are not known from the observation, the entropy production that is statistically obtained does not explicitly contain the information of the demon, and thus it is denoted by unconditional entropy production $\sigma_{X}$. These two different observations produce proper entropy productions, both of which satisfy the fluctuation theorem, regardless of whether or not a demon is involved. The difference between them indicates a portion of the energy dissipation caused by the demon's manipulation, i.e.,
\begin{equation}
\sigma_{I}=\sigma_{X|Y}-\sigma_{X},
\label{EqsigI}
\end{equation}
which characterizes a different form of energy dissipation from the entropy production, called the dissipative information~\cite{jsmte16-p04010,rmp91-045004,sa7-eabf1807,prxq_zhang,qft_zhang}. Hence,  to fully describe the thermodynamics of a controlled system, we need to know their own fluctuation theorems (including the SUT)~\cite{sa7-eabf1807,prxq_zhang,qft_zhang}, i.e.,
\begin{equation}
\langle e^{-\sigma_{X|Y}}\rangle=1 (\text{SUT}), \ \langle e^{-\sigma_{X}}\rangle=1, \ \langle e^{-\sigma_{I}}\rangle=1,
\label{EqFTs}
\end{equation}
with $\langle\sigma_{X|Y}\rangle\ge0$, $\langle\sigma_{X}\rangle\ge0$, and $\langle\sigma_{I}\rangle\ge 0$. Combining them leads to the actual total dissipation $\langle\sigma_{X|Y}\rangle$ of the controlled system, satisfying a nontrivial lower bound $\langle\sigma_{X|Y}\rangle \ge\langle\sigma_{I}\rangle$~\cite{sa7-eabf1807,prxq_zhang}, which is  tighter than the lower bound $\langle\sigma_{X|Y}\rangle \ge 0$ given by SUT.
%Besides, the SUT gives a lower bound of $0$ for the conditional entropy production $\langle\sigma_{X|Y}\rangle$, but this lower bound is achieved when the demon slowly operates the system in a quasi-static state, and is therefore of limited practical value. The demon can effectively control the system, accompanied by a non-zero dissipative information, $\langle\sigma_{I}\rangle>0$, which is the intrinsic energy dissipation due to the demon involved, and works as a nontrivial tighter lower bound of the entropy production~\cite{sa7-eabf1807,prxq_zhang},
%\begin{equation}
%\langle\sigma_{X|Y}\rangle \ge\langle\sigma_{I}\rangle.
%\label{EqLB}
%\end{equation}

To demonstrate the fluctuation theorems, we need to experimentally obtain the dissipative entities in Eq.~(\ref{EqsigI}) by the relations~\cite{SM}
\begin{equation}
\sigma_{X|Y}=\ln\frac{q_{x_c|y}}{p^{eq}_{x_c}},\sigma_{X}=\ln\frac{p_{x_c}}{p^{eq}_{x_c}},\sigma_{I}=\ln\frac{q_{x_c|y}}{p_{x_c}},
\label{EqDissipative}
\end{equation}
where $p^{eq}_{x_c}$ is given by the above Boltzmann distribution, and the conditional probability ($q_{x_c|y}$) and unconditional probability ($p_{x_c}=\sum_y p(y)q_{x_c|y}$) are obtained by the Boltzmann distribution with the error probability $p(y)$, and the mapping from the initial state $x_0$ to the controlled state $x_c$, respectively.

\begin{figure}[htpb]
 	\centering
\includegraphics[width=8.5cm,height=7.1cm]{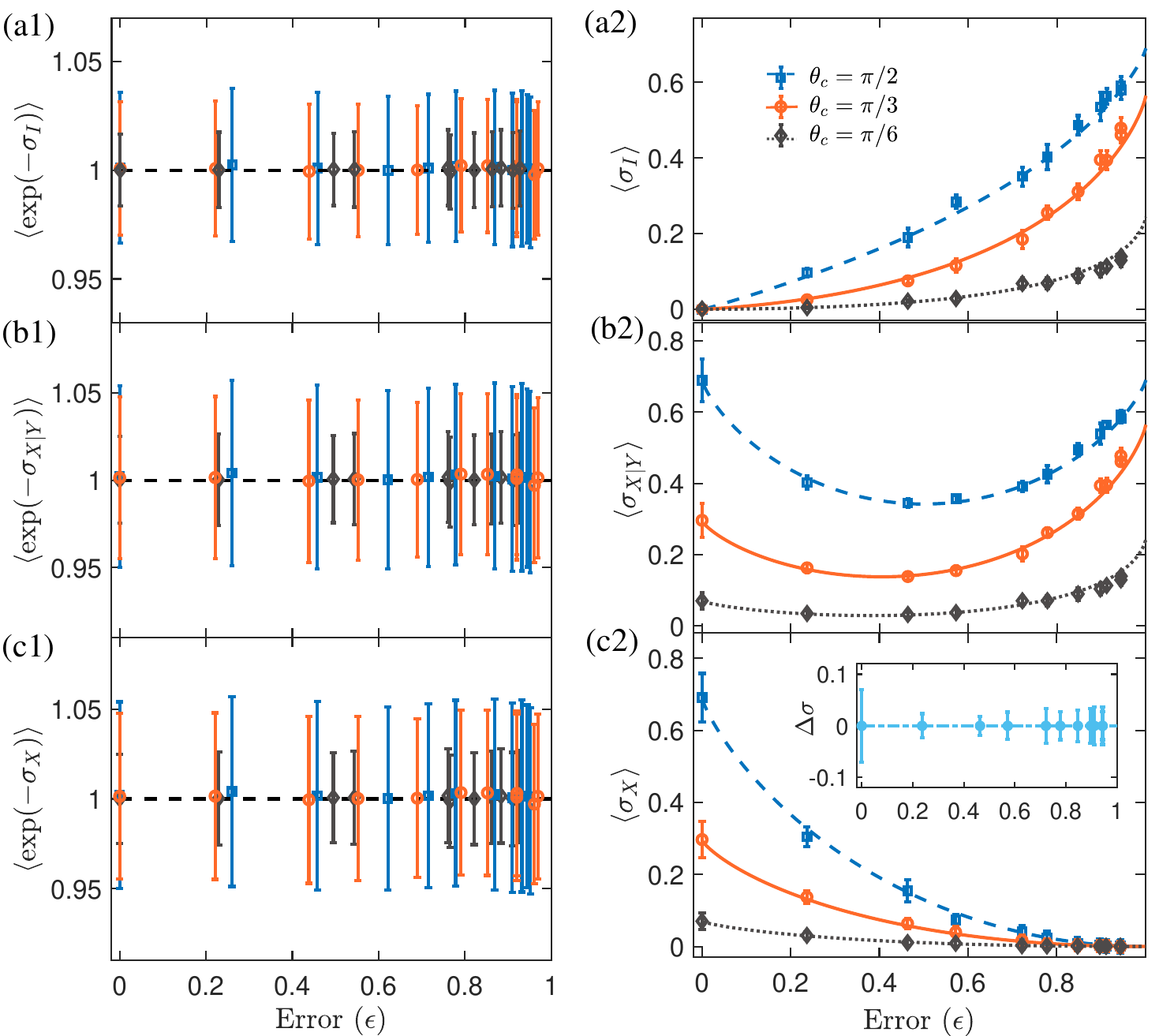}
\caption{The fluctuation theorems (a1-c1) and ensemble average (a2-c2) of the dissipative information $\sigma_I$, the conditional entropy production $\sigma_{X|Y}$  and the unconditional entropy production $\sigma_{X}$ for different initial mixed states, prepared by laser irradiation with durations $\theta_c=\pi/2,~\pi/3$ and $\pi/6$, respectively. Their entropy production difference $\Delta \sigma:=\langle\sigma_{X|Y}\rangle-\langle\sigma_X\rangle-\langle\sigma_I\rangle$, see the inset of (c2). The dot data and curves (also in the following figures) represent the experimental observation and analytical calculation, respectively. }
\label{Fig2a}
\end{figure}

Figure~\ref{Fig1}(b) presents the above procedures implemented in a $^{40}$Ca$^+$ ion confined in a linear Paul trap. To generate a Boltzmann distribution in the qubit levels, we first prepare a superposition by a carefully calibrated 729-nm laser pulse with $\theta_c=\Omega_1\tau_1$ and $\tau_1$ being the irradiation duration of the laser. Then, we wait for a period of time (about 2 ms for thermalization) during which the superposition state dephases, resulting in a mixed state with population in $\mid\downarrow\rangle$ state as $p_\downarrow=\frac{1}{2}(1+\cos\theta_c)$. This gives an effective inverse temperature of the system as $\beta=\ln [(1+\cos\theta_c)/(1-\cos\theta_c)]$, such as, the effective temperature $T=\beta^{-1}=440$, $0.94$ and $0.38$ for $\theta_c=\pi/2$, $\pi/3$ and $\pi/2$ \cite{EXP}, respectively. Besides, we introduce a controllable measurement error $y\neq x_0$ with the probability $\epsilon=p(y\neq x_0|x_0)=1-\exp(-\zeta\theta)$ and the pulse length $\theta = \Omega_2\tau_2$ by controlling the duration $\tau_2$ of the 854-nm laser. In Fig.~\ref{Fig1}(d), the decay parameter is experimentally measured as $\zeta=1.94(1)$ (the data in the parenthesis denoting the measurement uncertainty). By adjusting the duration of the laser irradiation $\theta$, we can vary the measurement error probability $\epsilon$ from 0 to 1, and acquire the joint probabilities and conditional probabilities of the system and demon in Fig.~\ref{Fig1}(e) and (f)~\cite{SM}, respectively.

To implement the feedback control, the experimenters take the role of the demon by performing control operations on the ion according to the results of the noisy measurements: If the ion is measured to be in the higher level, the demon couples the system to a battery (storing work) through a red sideband pulse of the 729-nm laser governed by the Hamiltonian $H_{r}=\hbar\tilde{\eta}\Omega_1(a\sigma_+ + a^{\dagger}\sigma_-)$, with Lamb-Dicke parameter $ \tilde{\eta}=0.11 $ and duration $\tau_r=\pi/\tilde{\eta}\Omega_1$, which conserves the quantum number. If the battery, simulated by the $z$-axis vibrational mode of the ion, is initialized in ground state $|0\rangle$, only the translation $\mid\uparrow\rangle |0\rangle\mapsto\mid\downarrow\rangle |1\rangle$ occurs during the control process of the demon, i.e., the demon extracts a nonzero work only when the system is actually in the higher level. The extracted work is quantified by the increase of the phonon number stored in the battery.

As shown in Fig.~\ref{Fig2a}, our experimental results fit the theoretical predictions well with the error bars less than 6$\%$ for all the three dissipative entities. This presents strong experimental evidences that both the fluctuation theorems and the nonnegativity of the ensemble average of the dissipative entities hold. Therefore, the system does not violate the second law of thermodynamics under the Szilard engine model controlled by the demon, consistent with the fluctuation theorems of the proper entropy productions $\sigma_{X|Y}$ and $\sigma_{X}$.  Importantly, the nonnegativity of the average of the dissipative information shows that $\langle\sigma_I\rangle$ is a novel kind of dissipative entity that is distinct from the entropy production. Besides, the equality relation of the summation $\sigma_{X|Y}=\sigma_X+\sigma_I$ can be observed from the inset of Fig.~\ref{Fig2a}(c2), and the tighter lower bound $\langle\sigma_{X|Y}\rangle \ge\langle\sigma_{I}\rangle$ of the entropy production is also witnessed in Fig.~\ref{Fig2a}(c2), together with the relation in Eq.~(\ref{EqFTs}).

\begin{figure}[htpb]
 	\centering
\includegraphics[width=8.5cm,height=6.8cm]{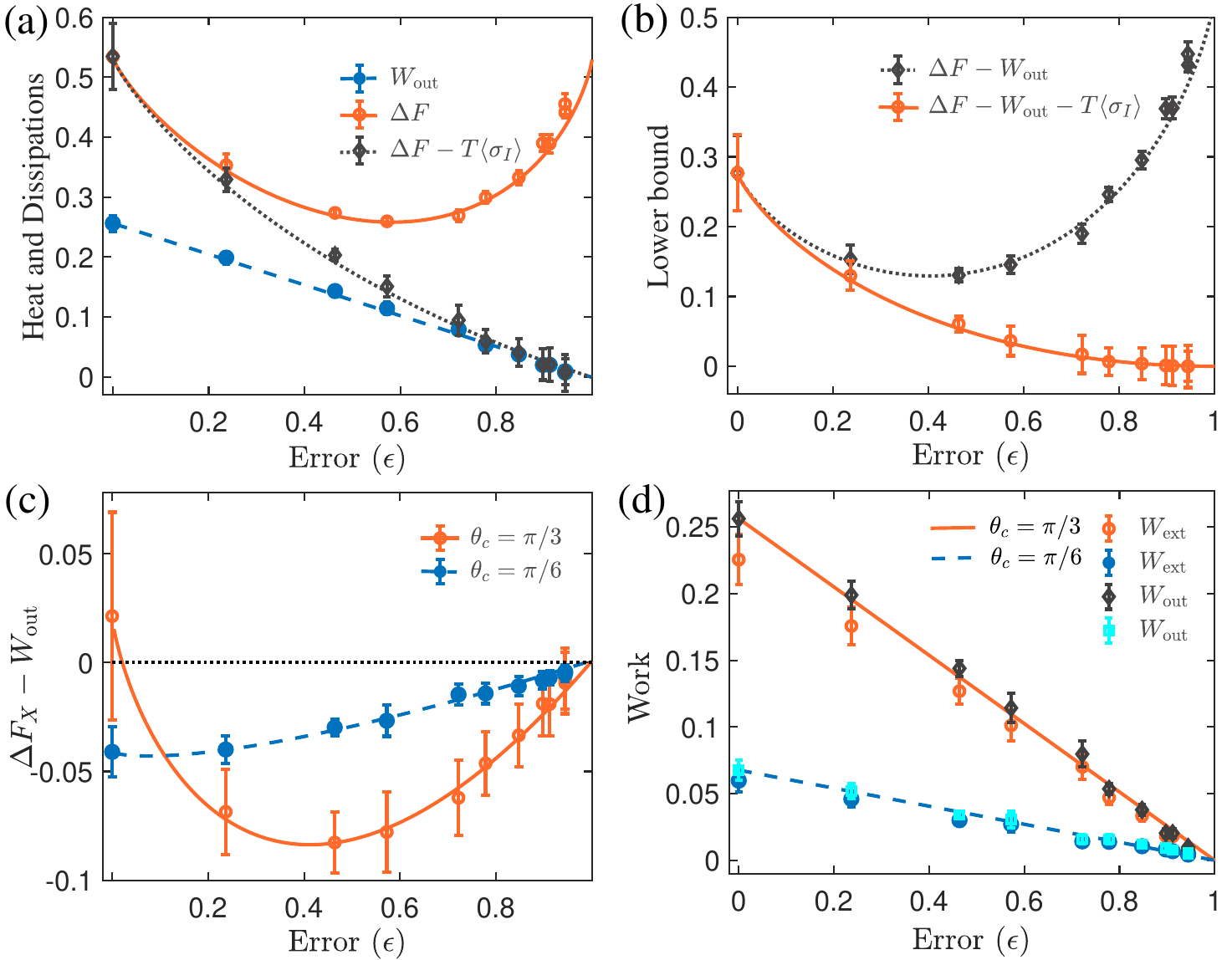}
\caption{The thermodynamic second law inequalities of work (a) for the conventional and tighter form, and (b) the conventional and tighter low bound, under $\theta_c=\pi/3$. (c) The coarse-grained version of SUT for different mixed states. (d) The output work $W_{\rm out}$ and stored work $ W_{\rm ext}$ in the vibrational mode after the control process.}
\label{Fig2b}
\end{figure}

To see the significant impact of the dissipative information on the thermodynamics of the system, we experimentally demonstrate the inequalities of thermodynamic second law and the tighter bound of the entropy production $\langle\sigma_{X|Y}\rangle$ given by the dissipative information, in comparison to the bound given by the SUT.
%As discussed earlier, the SUT is the fluctuation theorem for the total entropy production $\sigma_{X|Y}$ (i.e., the first equation in Eq.~(\ref{EqFTs})) and leads to the conventional second law of thermodynamics, i.e., $\langle\sigma_{X|Y}\rangle\ge 0$.
In particular, the total entropy production can be written in terms of the average of the extracted work and the free energy difference in our measurement-controlled experiment as $\langle\sigma_{X|Y}\rangle=\Delta F-W_{\rm out}$, where we define the extracted work and free energy difference as $W_{\rm out}=-\langle W_{X|Y}\rangle$ and $\Delta F=-\langle F_{X|Y}\rangle$~\cite{SM}. Thus, the SUT yields an upper bound of the extracted work $W_{\rm out}$, given by the free energy difference $\Delta F$~\cite{SU1,SU2},
\begin{equation}
W_{\rm out}\le \Delta F,
\label{EqConventionalbound}
\end{equation}
which also implies the upper bound inequality of the $W_{\rm out}$ as $\langle Q_{X|Y}\rangle\le T \langle \Delta S_{X|Y}\rangle$ with $ \Delta S_{X|Y}$ describing the stochastic entropy product~\cite{SM}. The free energy difference during the control contains the system information from the measurement. On the other hand, the dissipative information  $\langle\sigma_{I}\rangle$ quantifies the remaining information of the system right after the control. The mutual information between the controlled state $x_c$ and the measurement result $y$ is thus not used to extract work, but instead dissipates into the environmental bath. Therefore, the demon can extract work not higher than the difference between the free energy difference and the dissipative information, yielding a tighter upper bound of the extracted work $W_{\rm out}$ than the conventional bound in Eq.~(\ref{EqConventionalbound}), i.e.,
\begin{equation}
W_{\rm out}\le \Delta F-T\langle\sigma_{I}\rangle\le \Delta F.
\label{EqNewbound}
\end{equation}
Since the internal energy of the system remains unchanged, we can further obtain $\Delta F=T\langle\Delta S_{X|Y}\rangle$ and $W_{\rm out}=\langle Q_{X|Y}\rangle$~\cite{SM}. These relations have been experimentally witnessed in Fig.~\ref{Fig2b}(a-b), and some more details of the different process can be found in the supplementary materials~\cite{SM}. However, if we consider the coarse-grained entropy change $\Delta S_{X}=\langle\Delta S_{X|Y}\rangle_{X}$ with average only on the demon $Y$, the inequality of thermodynamic second law gives the coarse-grained version as $\langle Q_{X|Y}\rangle\leq T\langle\Delta S_{X}\rangle$, i.e., $W_{\rm out}\leq \Delta F_{X}$, with the coarse-grained free energy change $\Delta F_{X}=T\langle\Delta S_{X}\rangle$.  Figure~\ref{Fig2b}(c) presents the violation of the coarse-grained version in almost the whole error range, illustrating that the improper entropy productions decompose the demon's control from the total entropy production.

\begin{figure}[t]
 	\centering
\includegraphics[width=8.5cm,height=4.5cm]{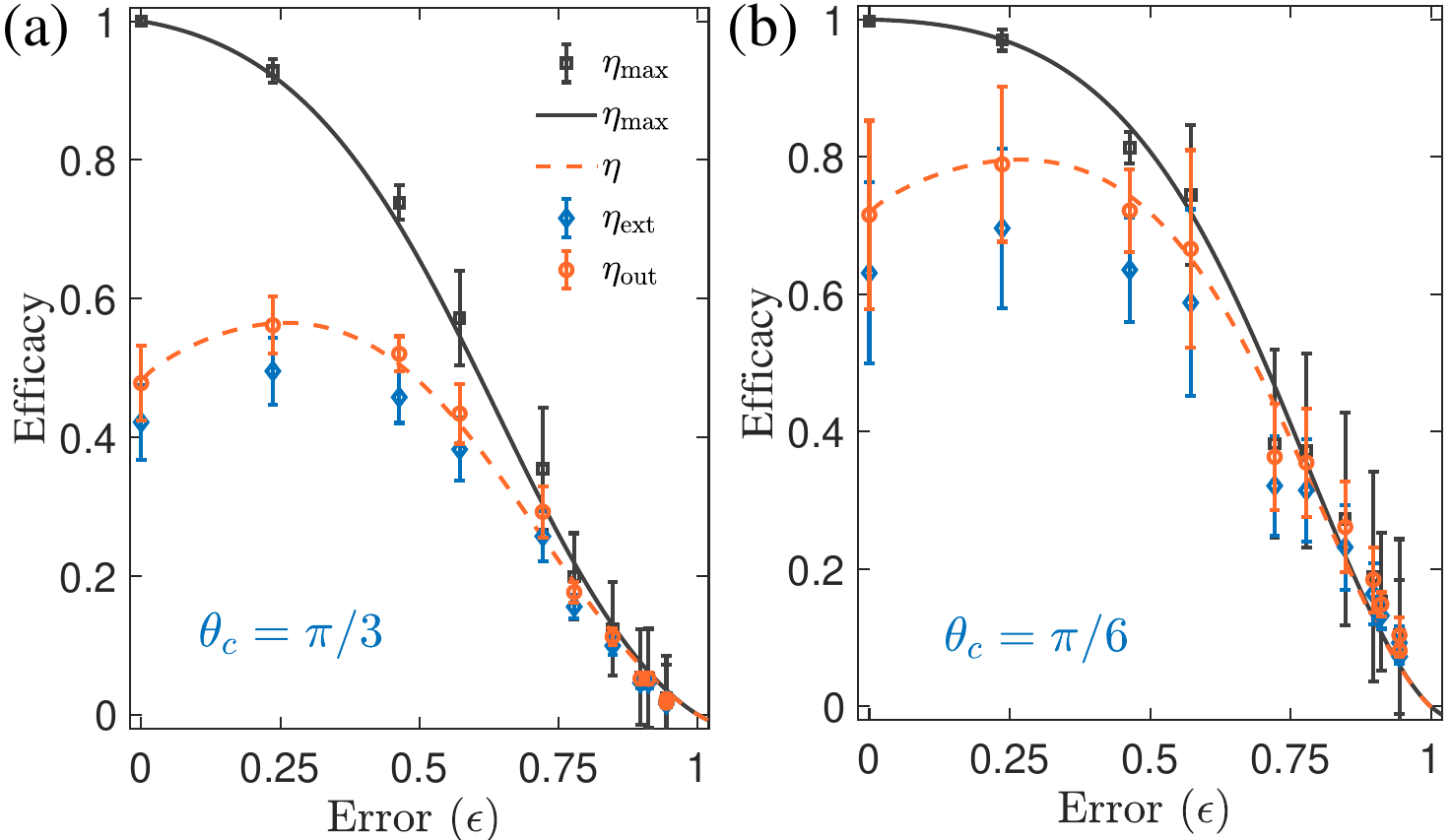}
\caption{The output work efficacy $\eta_{\rm out}$ (red data), actual efficacy $\eta_{\rm ext}$ (blue data), and the best efficacy $\eta_{\text{max}}$ (black data) of the demon in the whole control process. }
\label{Fig2c}
\end{figure}

The conventional bound of work in Eq.~(\ref{EqConventionalbound}) suggests that the control efficiency of the demon, or the demon's efficacy in work extraction, can be quantified by the ratio of the extracted work to the free energy difference, i.e., $\eta=W_{\rm out}/\Delta F$. According to the SUT, the efficacy of the demon should not be higher than 1 (or 100\%), i.e., $\eta\le 1$. However, the efficacy of 1 can only be achieved at quasi-static limit, which is of no practical significance. The tighter bound in Eq.~(\ref{EqNewbound}) indicates that the demon's efficacy is limited by the dissipative information. This suggests that the maximum achievable efficacy is not 1, but rather the difference between the free energy difference and the dissipative information~\cite{sa7-eabf1807,SM}, i.e.,
%\begin{equation}
$\eta_{\rm ext}  \leq \eta_{\rm out} \leq \eta_{\text{max}} \leq 1$,
%\label{EqEfficacy}
%\end{equation}
where the ideal output work efficacy $\eta_{\rm out}=W_{\rm out}/\Delta F$ describes the efficacy without considering the dissipation in the output process, the best efficacy is $\eta_{\text{max}}=1-T\langle \sigma_I\rangle/\Delta F$, and we also add an actual efficacy obtained by the extracted work stored in the auxiliary system $\eta_{\rm ext}=W_{\rm ext}/\Delta F$. For the Szilard engine control protocol, as shown in Fig.~\ref{Fig2b}(d), we always have the positive work outputs $W_{\rm out}=(1-\epsilon)p_{\uparrow}E$ for any error. Figure~\ref{Fig2c} demonstrates experimental verification of the relations and inequalities of these efficacy. The actual output work process executed by the red-sideband transition is affected by the dissipative elements of the vibrational mode, such as the imperfect cooling. Particularly, the initial average phonon number of the vibrational mode is $0.14(5)$ after cooling, and the final average phonon number is measured as $1.02(3)$ after implementing the control protocol, which gives an increase of the average phonon number $\Delta n=0.88(5)$, i.e., the $88\%$ storage efficiency, thus leading the actual efficiency $\eta_{\rm ext}$ to be deviated from the analytical curves and ideal output work efficacy $\eta_{\rm out}$.

In summary, our experiment has demonstrated the first verification of the recently proposed set of fluctuation theorems, confirming the proposed fluctuation theorems and observing tighter bounds on the extracted work and control efficacy of the demon than the SUT predicted limits. To further demonstrate the validity of demon-involved fluctuation theorems, we have also carried out an experiment based on a more popular state flipping control protocol, see \cite{SM}. We believe that the presented experimental results at the atomic level would be helpful for the future design of optimal control, and further understanding thermodynamic characteristics in the quantum regime, particularly for the characterization of Maxwell's demon in quantum information processing \cite{explain}.

This work is supported by the National Key Research and Development Program of China under Grant No. 2022YFA1404500, by Cross-disciplinary Innovative Research Group Project of Henan Province under Grant No. 232300421004, National Natural Science Foundation of China under Grant Nos. 1232410, U21A20434, 12074346, 12274376, 12074232, 12125406, 12374466, 12074390, 92265107, 12305028, 12247103, 12234019, by Natural Science Foundation of Henan Province under Grant Nos. 232300421075, 242300421212, by Major science and technology project of Henan Province under Grant No. 221100210400, by K. C. Wong Education Foundation (GJTD-2019-15), and by Nansha senior leading talent team technology project under Grant No. 2021CXTD02.


\begin{thebibliography}{99}
\bibitem{EXP} This Effective temperature, beyond its classical meaning, is a key indicator of control efficiency. This is because the Szilard engine can only extract work when the system is in the higher energy level. If the effective temperature is too low, it is difficult for the system to be at a higher energy level, so it is difficult to extract work from the system, reducing the control efficiency.
\bibitem{explain} In the quantum regime, we notice some recent studies in the framework of the two-point measurement scheme, uncovering that the dissipative quantum correlation (quantum dissipative information) also follows the fluctuation theorem, which characterizes the additional work waste in the quantum Maxwell's demon. If the constraints of the two-point measurement scheme are relaxed (e.g., by projections on both the initial and final states), the quantum dissipative information can be described by the quasiprobability fluctuation theorem, which statistically captures quantum data-processing inequality. Therefore, further exploring quantum fluctuation theorems of quantum dissipative information is highly expected in the future.
\end{thebibliography}
\end{document}